\documentclass[11pt]{article}

\usepackage{amssymb}
\usepackage{amsmath}
\usepackage{xcolor}

\begin{document}

\title{\hfill {\small \medskip }\\
\textbf{Constant curvature black holes in Einstein AdS gravity: Euclidean action and thermodynamics}}
\author{Pablo Guilleminot$^{a}$, Rodrigo Olea$^{a}$ and Alexander N. Petrov$^{b}$\smallskip \\
$^{a}${\small \emph{Departamento de Ciencias F\'{\i}sicas, Universidad
Andres Bello,}}\\
{\small \emph{Santiago, Chile. \smallskip }}\\
$^{b}$\emph{\small Moscow MV Lomonosov State University, Sternberg
Astronomical Institute, }\\
\emph{\small Universitetskii Prospekt 13, Moscow 119992, Russia.}{\small \emph{%
\smallskip }}\\
{\small \texttt{yemheno@gmail.com, rodrigo.olea@unab.cl, alex.petrov55@gmail.com}}}
\date{}
\maketitle


\begin{abstract}
We compute the Euclidean action for constant curvature black holes (CCBHs), as an attempt to associate thermodynamic quantities to these solutions of Einstein anti-de Sitter (AdS) gravity. CCBHs are gravitational configurations obtained by identifications
along isometries of a $D$-dimensional globally AdS space, such that the Riemann tensor remains constant. Here, these solutions are interpreted as extended objects, which contain a $(D-2)$-dimensional de-Sitter brane as a subspace. Nevertheless, the computation of the free energy for these solutions shows that they do not obey standard thermodynamic relations.
\end{abstract}

\section{Introduction}

Since the proposal of the four laws of black hole thermodynamics \cite{BCH}, black holes are considered as objects
with entropy, mass, angular momentum and electric charge as extensive quantities. According to the no-hair theorem, any black hole solution is completely determined by the last three. Moreover, despite the fact that gravity is not consistent yet with quantum theory, thermal field theories emerge when performing a Wick rotation. The free energy $F=-T\ln Z$ (where $Z$ is the partition function) within the saddle point
approximation has a leading order given by the evaluation of classical Euclidean action. These properties, linking geometry to quantum behavior, represent a consistency check for new solutions. Surface gravity $\kappa$, defined in a Killing horizon, plays the role of temperature. The black hole temperature can be understood as the period of the Eucidean time in order to avoid conical singularities. In Einstein gravity, the horizon area is proportional to entropy according to the Bekenstein-Hawking area formula \cite{Bek,Haw}.

Furthermore, Schwarzschild black holes have negative specific heat
$C_P=T\frac{\partial S}{\partial T}=\frac{\partial M}{\partial T}$ in flat spaces \cite{Haw}, where the pressure $P$ is related to the cosmological constant in a extended thermodynamic phase space \cite{GKB}. Negative specific heat leads to a violation of the laws of thermodynamics when the system is on a thermal bath.
To tackle this problem, York proposed to put the black hole in a small cavity \cite{York}. In this reference, it is shown that for a sufficiently small cavity the specific heat changes sign, defining a well-behaved canonical ensemble. Asymptotically AdS (AAdS) spaces naturally provide this setup such that the specific heat may change sign varying the thermodynamic variables of the system. Indeed, Hawking and Page proposed the existence of a phase transition between a black hole solution and thermal AdS \cite{HawPage}.

On the other hand, there is a relation between Noether's theorem and the computation of thermodynamic quantities. More precisely, the entropy can be worked out from a conserved current associated to a diffeomorphic isometry evaluated at the horizon \cite{Wald}. However, associating internal energy to this flux at radial infinity has a few subtleties. In particular, this quantity gives rise to a fraction of the Hamiltonian mass plus a divergent term. This issue can be fixed by a suitable choice of boundary terms, recovering Smarr-type relations  from a renormalized Euclidean AdS gravity action \cite{EJM,Pa-Sk,EGB-NED}.

In the last twenty five years, there has been a growing interest in gravitational solutions obtained from
global identifications of a locally AdS space, partly motivated by the introduction of the
Ba\~nados-Teitelboim-Zanelli (BTZ) black hole in $(2+1)$ dimensions \cite{BTZ}.

A proposal that extends this notion to higher dimensions was given by Ba\~nados in Ref.\cite{Banados}.
This comes as the natural generalization of the construction of the BTZ black hole as a quotient space of AdS$_3$
by a discrete group \cite{BHTZ}. This class of spacetimes drew considerable attention from the community for some time, especially in the context of anti de-Sitter/Conformal Field Theory (AdS/CFT) duality. Due to the appearance of two parameters in the metric, it was thought that they corresponded to mass and angular momentum. However, the interpretation of the parameters of the solution as conserved quantities has been recently disputed \cite{GOP}.

In the next lines, we study thermodynamic properties of CCBHs. The entropy and free energy are computed
from direct evaluation of the Einstein AdS
action, properly renormalized by the addition of extrinsic counterterms \cite{Olea2n,OleaKTs}.

\section{Construction of CCBHs} \label{Cons}

The obtention of (2+1)- dimensional black holes as  quotient spaces of global AdS$_3$ \cite{BHTZ}
raises the question of what other kinds of causal structures in 3D spacetimes are constructed using a subgroup of its symmetry group SO(2,2). Such a classification requires a criterion in order to tell among
the different resulting spaces.

In Euclidean 2D spaces, identifications are obtained by gluing edges, such that the resulting geometries may be orientable or non-orientable. Therefore, the surface are classified by the number of
holes and handles or crosscaps.

In the 3D case, the closest to a characterization of geometric structures is given by the  Thurston's geometrization conjecture \cite{Thurston}, based on the fact that 3-manifolds admit a canonical decomposition \cite{Perelman1,Perelman2}.

In turn, our interest here goes out of the scope of the above classification, as the subject of study here is the causal structure produced by identifications in maximally-symmetric spaces in three and higher dimensions.

Due to the topological character of (2+1)-dimensional gravity, it was thought that the theory could not have black hole solutions. As the Weyl tensor is zero, the curvature is fully determined by the Ricci tensor, what renders nontrivial the construction of gravitational objects.
Indeed, BTZ black hole is locally indistinguishable from pure AdS. Different global properties arise from a quotient space AdS$_3$/$\Gamma$, where $\Gamma$ is a discrete subgroup \cite{BHTZ,HolShe}. Points in the universal covering of AdS$_{3}$ then appear identified along isometries.

The type of singularity obtained in the causal structure changes with the isometry $\xi^{\mu}$ used for the identification. Indeed, both rotations and boosts may leave fixed points in the resulting space\footnote{In the covering space, the signature is $(-,+,+,-)$. Rotations refer to operations that involve coordinates with the same
signature, in opposition to boosts, which mix coordinates with opposite sign.}.
This fixed points give rise to singularities in the quotient space \cite{ABBHP} of conical-type for rotations \cite{MisZan} and BTZ-type for boosts \cite{BHTZ}. If these singularities are hidden behind an event horizon, they identifications give rise to black hole solutions \cite{CotGib}.

In order to distinguish between the different causal structures obtained by identifications, \cite{BHTZ}
one has to analyze the properties of the Killing field of SO(2,2) $\xi$, which defines the equivalence class $\Gamma=e^{\tau\xi}$, where $\tau$ is a finite parameter. The information needed is encoded in the relation $\xi=\frac{1}{2}W^{ab}J_{ab}=w^{ab}x_b\partial_a$, where  $W^{ab}$ and $J_{ab}=x_{b}\partial_{a}-x_{a}\partial_{b}$ are an antisymmetric matrix in $a,b$ and the generators for the SO(2,2) group, respectively. When $\xi$ acts on a coordinate $x^c$ as $\xi x^c=W^{c}_ax^a$, it defines an infinitesimal symmetry transformation
$x'^a=(\delta^a_b+\epsilon w^a_b)x^b$ \cite{HP}. Then, a comparison is made by obtaining the eigenvalues for each $W^{ab}$, what classifies the possible identifications \cite{BHTZ,HP}.

Finally, the causal structure for any of these spaces is analyzed, bestowing special attention
to how the point $\xi^\mu\xi_\mu=0$ is disconnected the sector where $\xi$ is spacelike,
which is accessible for an observer \cite{HP}. As identifications may bring in closed timelike
curves, the region where they appear ($(\xi^\mu\xi_\mu<0)$) must be cut off from the original AdS$_3$ space \cite{BHTZ}.

The procedure outlined above was carried out in $(2+1)$ dimensions in Ref.\cite{BHTZ} and, in a greater detail, in Ref.\cite{HP}. It was also extended to provide a partial classification in $(3+1)$ dimensions in the latter reference. A deeper analysis and further black hole solutions in 4D were
shown in Ref.\cite{ABBHP}. For higher dimensions, Ref.\cite{Banados} proposes the construction of a solution identifying
along one single boost isometry, following the steps in \cite{BHTZ}. Due to it properties they were dubbed as
constant-curvature black holes. CCBHs are constructed from identifications in a pure AdS space, which leaves invariant the  curvature
\begin{equation} \label{LAdS}
R_{\alpha \beta \mu \nu}=-\frac{1}{\ell^{2}}\left(g_{\alpha\mu}g_{\beta\nu}-g_{\alpha\nu}g_{\beta\mu}\right)\,.
\end{equation}
Taking the trace in a pair of indices in the above relation makes clear that a constant-curvature object solves the Einstein equations $R_{\mu\nu}=-(D-1)g_{\mu\nu}/\ell^{2}$. However, on the contrary to the three-dimensional case, it seems to be that identifications made in AdS vacuum cannot change the value of the global charges (mass and angular momentum) \cite{GOP}.

\section{CCBHs as a brane}

Let us consider the general ansatz for the metric describing an extended object
\begin{equation} \label{ansatz}
ds^2=A^2(r)\omega_{ab}dy^ady^b+B^2(r)dr^2+C^2(r)\gamma_{mn}dx^mdx^n\,,
\end{equation}
which is a warped product between a Lorentzian subspace with metric $\omega_{ab}$ and another
Euclidean one with metric $\gamma_{nm}$. The indices ${a,b}$ denote the coordinates on the brane, whereas ${n,m}$ are
stand for those in the transversal section

In order to make contact with existing solutions in the literature, we set $C(r)=r$ in the expressions for the curvature Eqs.(\ref{34_2}), (\ref{34_3}) and (\ref{34_5}), what leads to
\begin{eqnarray}
R^{rn}_{rm} & = & -\frac{1}{rB(r)}\left ( \frac{1}{B(r)}\right )^{\prime}=\frac{B^\prime(r)}{rB^3(r)}\delta_m^n\,, \label{ries_1} \\
R_{bm}^{an} & = & -\frac{A^\prime(r)}{rA(r)B^2(r)}\delta^{[bn]}_{[am]}\,, \label{ries_2} \\
R_{nm}^{pq} & = & \frac{1}{r^2}\mathcal{R}_{nm}^{pq}(\gamma)-\left(\frac{1}{rB(r)}\right )^2\delta^{[pq]}_{[nm]}\label{ries_3}\,,
\end{eqnarray}
where a prime stands for radial partial derivative.
We then impose the constant-curvature condition (\ref{LAdS}) in this general metric. From Eq.(\ref{ries_1}), it follows that
\begin{equation}
B^2(r) =  \frac{1}{\left (-\beta+\frac{r^2}{\ell^2}\right )}\,,
\end{equation}
where $\beta$ is an integration constant. This result, when plugged in Eq.(\ref{ries_2}), produces
\begin{equation}
A^2(r)  =  \alpha\left (-\beta+\frac{r^2}{\ell^2}\right )\,,
\end{equation}
where $\alpha$ is a second integration constant, which can be absorbed in a redefinition of coordinates on the brane.
As a consequence, we work with a single function in the metric, of the form
\begin{equation}\label{fAB}
f^2(r)\equiv A^2(r) = B^{-2}(r) = -\beta+\frac{r^2}{\ell^2}.
\end{equation}
Taking into account the form of Eq.(\ref{fAB}), the remaining components the spacetime Riemann tensor (\ref{34_4}) and (\ref{34_5}) for the present case are
\begin{equation}\label{RB}
 \mathcal{R}^{ab}_{cd}(\omega) =  f^2(r)\left ( -\frac{1}{\ell^2} +\left (f^{\prime}(r)\right )^2\right )\delta^{[ab]}_{[cd]}
 =\frac{\beta}{\ell^2}\,\delta^{[ab]}_{[cd]}\,,
\end{equation}
and
\begin{equation}\label{RT}
\mathcal{R}_{nm}^{pq}(\gamma) = r^2\left(-\frac{1}{\ell^2}+\frac{f^2(r)}{r^2}\right )\delta^{[pq]}_{[nm]}
= -\beta\delta^{[pq]}_{[nm]}\,.
\end{equation}
We notice that, depending on the sign of $\beta$, there are three different topologies.
For $\beta<0$, the transversal section is a sphere and the brane is a Lorentzian space with
negative curvature. Global AdS in standard Schwarzschild-like coordinates can be obtained picking up a 0-brane.
On the other hand, it is straightforward to check that AdS spacetime in Poincar\'e coordinates is recovered for
$\beta=0$.

In the case $\beta$ is positive, the brane is a space of positive curvature and Lorentzian signature, i.e., a dS spacetime.
In order this to be consistent to Eq.(\ref{RT}), the transversal section must be an Euclidean space
of negative curvature, what is clearly impossible. Said this, we choose
$\omega_{ab}dy^ady^b=\frac{\ell^2}{\beta}d\Omega_{D-2}^2$ to satisfy Eq.(\ref{RB}). Thus, the only
 possibility is to have a constant-curvature solution is to consider a $(D-3)$-brane, such that there is a single angular coordinate $\gamma_{mn}dx^mdx^n=d\phi^2$.
With all the above procedure, and absorbing once again the constant with a redefinition of coordinates, we can reproduce the metric of CCBHs
\begin{equation}\label{CCBH}
ds^2=f^2(r)d\Omega_{D-2}^2+\frac{dr^2}{f^2(r)}+r^2d\phi^2\,,
\end{equation}
 with the particular value $\beta=\frac{r_+^2}{\ell^2}$, that is,
 \begin{equation}\label{fCCBH}
f^2(r) = \frac{r^2-r_{+}^2}{\ell^2}\,.
\end{equation}

In the Ref.\cite{Banados}, the parameter $r_{+}$ is thought as associated to the mass of the solution, such that this object
is not globally equivalent to AdS vacuum. The metric has a very similar form to the one of BTZ black hole in three
dimensions, but the interpretation of $r=r_{+}$ as a horizon is far more elusive.
In particular, the fact that the interior region cannot be even defined in this parametrization prevents a further analysis of the causal structure.\\
The identifications that explicitly realize this solution are performed in a $D$-dimensional AdS space, which is taken as a hypersurface defined in $%
(D+1)$-dimensional $\mathbb{R}^{\mathrm{2,D-1}}$ space, subjected to the constraint
\begin{equation}
-x_{0}^{2}+x_{1}^{2}+\cdots +x_{D-1}^{2}-x_{D}^{2}=-\ell ^{2}.
\label{AdSUC}
\end{equation}
Using the following set of variables
\begin{eqnarray} \label{iden1}
x_\alpha & = & \ell\sinh\rho\, y_\alpha\,, \notag \\
x_{D-1} & = & \ell\cosh\rho \sinh\left ( \frac{r_+\phi}{\ell}\right )\,, \notag \\
x_D & = & \ell\cosh\rho \cosh\left ( \frac{r_+\phi}{\ell}\right )\,,
\end{eqnarray}%
where $\alpha=0,...,D-2$. The condition $\eta_{\alpha\beta}y^\alpha y^\beta=1$ defines a dS subspacetime.
The resulting metric takes the form
\begin{equation} \label{CCBHclean}
ds^2=\ell^2(\sinh^2\rho\, d\Omega_{D-2}^2+\frac{r_+^2}{\ell^2}\cosh^2\rho\,d\phi^2+d\rho^2)\,.
\end{equation}
This metric can be cast in the original form (\ref{CCBH}) given in Ref.\cite{Banados} by setting $r=r_+ \cosh \rho$.
Ensuring that $\phi$ is a compact variable, by the identification $\phi \sim \phi+2\pi $, one finds the CCBH geometry.

In what follows, we study the thermodynamic properties of CCBHs, obtained from the corresponding Hemholtz free energy, computed directly from the Euclidean action.

\section{Free energy for CCBHs}

Thermodynamics for black holes is derived from the Euclidean action $I_{E}$, as it is proportional
to the free energy of the system. In Einstein AdS gravity, the Euclidean action is proportional to the volume 
element of the spacetime. Therefore, in order to obtain finite physical quantities one needs to rely on background-subtraction methods or a given renormalization scheme. Background-subtraction procedures to regulate the $I_E$ consider the difference between the Euclidean action evaluated for the solution of interest and the Euclidean action for the corresponding background. This, in general, requires matching conditions in the Euclidean period in the asymptotic region, what means that a vacuum state is endowed with thermal properties \cite{HawPage}.

However, in the CCBH solution, the limit $r_+\rightarrow 0$ does not recover global AdS \cite{Banados,GOP},
what renders the latter space unsuitable as a background.
 In turn, standard background-independent methods, consider the addition of counterterms as surface terms in the gravity action. This is especially appropriate to deal with the general problem of extracting holographic information (e.g., Weyl anomaly) from AdS gravity, in the context of gauge/gravity duality \cite{EJM,BK}.
 
The main drawback of the above prescription is the lack of a closed formula for the counterterm series in an arbitrary dimensions.
A proposal for counterterms -of a different sort- which produces a finite Euclidean action for AAdS solutions in any dimension, in background-independent fashion was given in Refs.\cite{Olea2n,OleaKTs}. 

\subsection{Even dimensions}

The addition of the Kounterterm series to the even-dimensional
Einstein-Hilbert action with negative cosmological constant is equivalent,
by virtue of the Euler theorem in $D=2n$ dimensions, to a topological invariant \cite{Olea2n}
\begin{equation}
I_{ren}^{2n}=-\frac{1}{16\pi G}\int\limits_\mathcal{M} d^{2n}x\sqrt{-g}\big[R-2\Lambda -\frac{(-\ell
^{2})^{n-1}}{2^{n}n(2n-2)!}\,\delta _{\left[ \mu _{1}\cdots \mu _{2n}\right]
}^{\left[ \nu _{1}\cdots \nu _{2n}\right] }R_{\nu _{1}\nu _{2}}^{\mu _{1}\mu
_{2}}\cdots R_{\nu _{2n-1}\nu _{2n}}^{\mu _{2n-1}\mu _{2n}}\big]\,,
\label{Ireg2n}
\end{equation}%
with a fixed coupling given in terms of the AdS radius. The only difference is
that the action (\ref{Ireg2n}) is shifted by a constant proportional to the
Euler characteristic of the manifold. It was shown in Ref.\cite%
{Miskovic-Olea4D} that this mechanism of regularization for even-dimensional
AdS gravity action is equivalent to the Holographic Renormalization program.

The regularized action may be written in terms of the Riemman tensor and the
antisymmetric Kronecker delta of rank $2$ in the form
\begin{eqnarray} \label{EHC}
I_{ren}^{2n} &=&-\frac{1}{2^{n+4}\pi G(2n-2)!}\int\limits_\mathcal{M} d^{2n}x\sqrt{-g}\delta _{\left[
\mu _{1}\cdots \mu _{2n}\right] }^{\left[ \nu _{1}\cdots \nu _{2n}\right] }%
\big[R_{\nu _{1}\nu _{2}}^{\mu _{1}\mu _{2}}\delta _{\left[ \nu _{3}\nu _{4}%
\right] }^{\left[ \mu _{3}\mu _{4}\right] }\cdots \delta _{\left[ \nu
_{2n-1}\nu _{2n}\right] }^{\left[ \mu _{2n-1}\mu _{2n}\right] } \notag \\
&&+\frac{(n-1)}{%
n\ell ^{2}}\delta _{\left[ \nu _{1}\nu _{2}\right] }^{\left[ \mu _{1}\mu _{2}%
\right] }\cdots \delta _{\left[ \nu _{2n-1}\nu _{2n}\right] }^{\left[ \mu
_{2n-1}\mu _{2n}\right] } +(-1)^{n}\frac{\ell ^{2n-2}}{n}\,R_{\nu _{1}\nu _{2}}^{\mu _{1}\mu
_{2}}\cdots R_{\nu _{2n-1}\nu _{2n}}^{\mu _{2n-1}\mu _{2n}}\big],
\end{eqnarray}
what is convenient for the discussion below.

The above expression can be schematically represented by the polynomial%
\begin{equation} \label{Iregspec}
I(x,y)=-\frac{\ell ^{2n-2}}{2^{n+4}\pi Gn(2n-2)!}\left(
(-1)^nx^{n}+n\,xy^{n-1}+(n-1)y^{n}\right) \,,
\end{equation}
where $x=R_{\mu\nu}^{\alpha\beta}$ and $y=\delta_{[\mu\nu]}^{[\alpha\beta]}/\ell^{2}$.
The key observation at this point is the fact that this polynomial
can be always factorized as
\begin{equation}
I(x,y)=-\frac{\ell^{2n-2}}{2^{n+4}\pi Gn(2n-2)!}(x+y)^2\sum_{k=0}^{n-2}(-1)^{n-2-k}(k+1)x^{n-2-k}y^k\,.
\label{factorization}
\end{equation}
Introducing a parametric integral, Eq.(\ref{factorization}) can be written as (See Appendix \ref{Beta})
\begin{equation}
 I(x,y)=-\frac{\ell^{2n-2}(n-1)}{2^{n+4}\pi G(2n-2)!}(x+y)^2\int\limits_0^1dt\,\left [y-t(x+y)\right ]^{n-2}\,.
\end{equation}
The previous representation allows to express the action as
\begin{equation} \label{GOPaction}
I_{ren}^{2n} = -\frac{\ell^{2n-2}(n-1)}{2^{n+4}\pi G(2n-2)!}\int\limits_{\mathcal{M}} d^{2n}x\sqrt{-g}\delta_{[\mu_1\cdots\mu_{2n}]}^{[\nu_1\cdots\nu_{2n}]}W^{\mu_1\mu_2}_{\nu_1\nu_2}W^{\mu_3\mu_4}_{\nu_3\nu_4}\int\limits_0^1 dt\,t\,\Xi^{\mu_5\mu_6}_{\nu_5\nu_6}(t)\times\cdots\times\Xi^{\mu_{2n-1}\mu_{2n}}_{\nu_{2n-1}\nu_{2n}}(t)\,,
\end{equation}
where
\begin{eqnarray}
\Xi^{\alpha\beta}_{\mu\nu}(t) & = & \frac{1}{\ell^2}\delta^{[\alpha\beta]}_{[\mu\nu]}-tW^{\alpha\beta}_{\mu\nu}\,,
\end{eqnarray}
and
\begin{equation}\label{OSW}
W^{\alpha\beta}_{\mu\nu}=R^{\alpha\beta}_{\mu\nu}+\frac{1}{\ell^2}\delta^{[\alpha\beta]}_{[\mu\nu]}\,,
\end{equation}
is the Weyl tensor for Einstein spaces. As a consequence, for CCBHs ($W^{\alpha\beta}_{\mu\nu}=0$
everywhere), the Euclidean action in $D=2n$ dimensions vanishes identically,
so does the free  energy $F = T I_{ren}$.

A well-defined notion of energy for black holes of any gravity theory is given in terms of the functional derivative
of the Lagrangian density $\mathcal{L}(g_{\mu\nu},R_{\alpha\beta\mu\nu})$ respect to the Riemann curvature, as discussed in Ref.\cite{Wald}.

The Noether-Wald charge is given by the expression
\begin{equation} \label{NWC}
Q[\xi]=\int\limits_{\Sigma} d^{2n-2}x\sqrt{-\sigma}\,n_\mu u_\nu E^{\mu\nu}_{\alpha\beta}\nabla^{\alpha}\xi^\beta\,,
\end{equation}
where $\xi^\beta$ is the Killing vector field associated to an isometry of the spacetime, $E^{\mu\nu}_{\alpha\beta} = \delta L/\delta R^{\alpha\beta}_{\mu\nu}$ and $\Sigma$ is a co-dimension 2 surface.
For the action (\ref{EHC}), $E^{\mu\nu}_{\alpha\beta}$ has the following expression:
\begin{equation}
E^{\mu\nu}_{\alpha\beta}\propto n\delta^{[\mu\nu\nu_1\cdots\nu_{2n-2}]}_{[\alpha\beta\mu_1\cdots\mu_{2n-2}]}\left [(-1)^n R_{\nu _{1}\nu _{2}}^{\mu _{1}\mu
_{2}}\cdots R_{\nu _{2n-3}\nu _{2n-2}}^{\mu _{2n-3}\mu _{2n-2}}
+ \frac{1}{\ell^{2n-2}}\delta _{\left[ \nu _{1}\nu _{2}\right] }^{\left[ \mu _{1}\mu _{2}\right] }\cdots \delta _{\left[ \nu
_{2n-3}\nu _{2n-2}\right] }^{\left[ \mu _{2n-3}\mu _{2n-2}\right] }\right ]\,.
\end{equation}
The previous formula  is always factorizable by
$R_{\nu _{1}\nu _{2}}^{\mu _{1}\mu_{2}}+\frac{1}{\ell^2}\delta _{\left[
\nu _{1}\nu _{2}\right] }^{\left[ \mu _{1}\mu _{2}\right] }$
anew. As a consequence, the integrand in (\ref{NWC}) is
proportional to the Weyl tensor (\ref{OSW}).
In other words, the Noether-Wald current is identically
zero for any constant curvature AdS solution. As the Noether-Wald charge is the entropy $S$
when evaluated at the horizon, the vanishing of the charges at infinity, i.e., $M=0$ and $J=0$
is consistent with the thermodynamic relation
\begin{equation} \label{thermo_relation}
F=M-TS-\Omega J\,,
\end{equation}
where $\Omega$ is the angular velocity of the horizon.

\subsection{Odd dimensions}

In $D=2n+1$ dimensions, the lack of an equivalent form for the Kounterterms
as a bulk term (i.e., a fully-covariant expression in the spacetime)
prevents a factorization of the action similar to Eq.(\ref{factorization}).

As a consequence, there is no alternative to the actual evaluation of the
Euclidean action for particular solutions of the Einstein equations.
To renormalize the action and work out the computations we pick a Gauss-normal frame
\begin{equation} \label{Gaussian}
ds^2=N^2(r)dr^2+h_{ij}(x,r)dx^idx^j\,.
\end{equation}
For this case, the action has the form
\begin{equation} \label{IEH}
I_{ren}^{2n+1}=-\frac{1}{16\pi G}\int\limits_\mathcal{M} d^{2n+1}x\sqrt{-g}\big[R-2\Lambda\big] + c_{2n}\int\limits_{\partial\mathcal{M}}d^{2n}x\,B_{2n}\,,
\end{equation}
where the surface term is
\begin{equation}
B_{2n}=2n\sqrt{-h}\int\limits_0^1dt\int\limits_0^tds\delta^{[j_1\cdots j_{2n}]}_{[i_1\cdots i_{2n}]}K^{i_1}_{j_1}
\delta^{i_2}_{j_2}\mathcal{F}^{i_3i_4}_{j_3j_4}\cdots\mathcal{F}^{i_{2n-1}i_{2n}}_{j_{2n-1}j_{2n}}\,.
\end{equation}
Here, $K_i^j$ is the extrinsic curvature defined as
\begin{equation}
K_{ij}=-\frac{1}{2N}\partial_r h_{ij}\,,
\end{equation}
and
\begin{equation}
 \mathcal{F}^{ij}_{kl}=\frac{1}{2}\mathcal{R}^{ij}_{kl}-t^2K^i_kK^j_l+\frac{s^2}{\ell^2}\delta^{i}_{k}\delta^{j}_{l}\,.
\end{equation}
The tensor $\mathcal{R}^{ij}_{kl}$ stands for the intrinsic curvature of the boundary metric $h_{ij}$ and $c_{2n}$ is the coupling constant \cite{OleaKTs}
\begin{equation}
 c_{2n}=\frac{1}{16\pi G}\frac{(-\ell^2)^{n-1}}{n(2n-1)!}\left [ \int_0^1dt(1-t^2)^{n-1} \right ]^{-1}\,
\end{equation}
that ensures that the variational principle is fulfilled.

The bulk part of the action (\ref{IEH}) can be rewritten as
\begin{equation}
I=-\frac{1}{16\pi G}\int\limits_{\mathcal{M}}d^{2n+1}x\,%
\sqrt{-g}\,\left (\frac{1}{2}\delta _{\lbrack \nu _{1}\nu _{2}]}^{[\mu _{1}\mu
_{2}]}R_{\mu _{1}\mu _{2}}^{\nu _{1}\nu _{2}}+\frac{(2n)(2n-1)}{\ell ^{2}}%
\right )\,,  \notag
\end{equation}

where Latin letters denote boundary components.
In the foliation (\ref{Gaussian}), the indices split as
$\mu =\left(r,i\right)$. Substituting in the general form of the action, it leads to
\begin{equation}
I=-\frac{1}{16\pi G}\int\limits_{%
\mathcal{M}}d^{2n}x\,dr\,\sqrt{-h}\,N\,\left [\frac{1}{2}\delta _{\lbrack
i_{1}i_{2}]}^{[j_{1}j_{2}]}\left( \,R_{j_{1}j_{2}}^{i_{1}i_{2}}%
\,+\frac{1}{\ell ^{2}}\,\delta _{[j_{1}j_{2}]}^{[i_{1}i_{2}]}\right )+
2\,R_{rj}^{ri}\delta _{i}^{j}\right ]\,.  \notag
\end{equation}%
One can recognize the component $(r,r)$ of the equations of motion
\begin{equation}
\mathcal{E}^\mu_\nu=\frac{1}{4}\delta^{[\mu\mu_1\mu_2]}_{[\nu\nu_1\nu_2]}\left (R^{\nu_1\nu_2}_{\mu_1\mu_2}+\frac{1}{\ell^2}\delta_{[\mu_1\mu_2]}^{[\nu_1\nu_2]}\right )\,,
\end{equation}
from the first two terms
and they vanish on-shell. Therefore, using Gauss-Codazzi relations, the action turn into
\begin{equation}
I=-\frac{1}{8\pi G}\int\limits_{\mathcal{M}}d^{2n}x\,
dr\,\sqrt{-h}\,NR_{ri}^{ri}=-\frac{1}{8\pi G}\int\limits_{\mathcal{M}}d^{2n}x\,
dr\,\sqrt{-h}\,\left [(K)'-NK^i_jK^j_i \right ]\,. \notag
\end{equation}
For the metric (\ref{CCBH})
the values of $K_j^i$ are
\begin{eqnarray}
K_a^b & = & -\frac{r}{\ell^2f(r)}\delta^a_b\,, \\
K_\phi^\phi & = & -\frac{f(r)}{r}\,,
\end{eqnarray}
where $a$,$b$ corresponds to the components of $\Omega$.

Let us consider a Wick rotation of the spacetime, where the time coordinate is $\tau=it$.
For CCBHs, any globally-defined set of coordinates will lack a timelike Killing vector
\footnote{This prevents computations of the entropy via microcanonical action as made in Ref.\cite{CM}}.
Because any version of the CCBH metric will be
explicitly time dependent, an Euclidean time period cannot be defined.
In doing so, the determinant of the boundary metric can be computed as
$\sqrt{h}=f^{D-2}(r)r\sqrt{\omega}$, where $\sqrt{\omega}$
is the determinant of the Euclideanized subspace $\Omega$. Plugging this in the last expression of the
action gives
\begin{equation}
I  =  -\frac{1}{4 G}\int dr\,rf^{2n-1}\left [
-\left ((2n-1)f'+\frac{f}{r}\right )'-(2n-1)\frac{f'^2}{f}-\frac{f}{r^2}\right ]
\int d\tau\,d^{2n-2}y\,\sqrt{\omega}\,. \notag
\end{equation}
Finally, the bulk Euclidean action can be written as a total derivative
\begin{equation}
I= \frac{1}{4 G}
\int dr\,\left (f^{2n}\right )'\int d\tau\,d^{2n-2}y\,\sqrt{\omega}\,,
\end{equation}
This expression evaluated in the horizon is zero, leaving behind
the integrand evaluated at radial infinity.

On the other hand, the boundary term $B_{2n}$ in Eq.(\ref{IEH}) can be substantially simplified
noticing that the only non-vanishing Riemann tensor of the boundary is $\mathcal{R}^{a_1a_2}_{b_1b_2}=\frac{\beta}{\ell^2f^2(r)}$, where $a,b$ indexes symbolize the components of $\omega$. Then, $B_{2n}$ can be rewritten as
\begin{eqnarray}
\int\limits_{\partial\mathcal{M}}d^{D-1}x\,B_{D}&=&4\pi n%
\frac{(2n-1)!}{(\ell^2)^{n-1}}\int_0^1dt\int_0^tds\left [ \left (f^2+\frac{r^2}{\ell^2}\right )\left ((1-t^2)\beta+(s^2-t^2)f^2\right )\right . \notag \\
&&\left . +(2n-2)\frac{r^2}{\ell^2}f^2(s^2-t^2)\right ]\left ((1-t^2)\beta+(s^2-t^2)f^2\right )^{n-2}\int d\tau\,d^{2n-2}y\,\sqrt{\omega}\,. \notag
\end{eqnarray}
The complete action (\ref{IEH}) reduces to a quantity of $\mathcal{O}(1)$ given by
\begin{equation} \label{EAF}
I=\frac{1}{4 G}\frac{(-1)^{n}(2n-1)!!^2}{(2n-1)!}\beta^n\int d\tau\,d^{2n-2}y\,\sqrt{\omega}\,,
\end{equation}
that is proportional to the vacuum energy density obtained in Ref.\cite{GOP}.
The above result makes manifest the fact that CCBHs do not follow standard thermodynamic relations in AAdS
gravity.

\section{Conclusions}

We have shown that CCBH solution can be understood as an extended object, i.e.,
a $(D-3)$-brane, where the surface $r=\beta\ell^2$ is not an event horizon
but simply the origin of the radial coordinate obtained by identifications of global AdS
spacetime.  \\
Using the metric obtained in Section \ref{Cons}, we derived
the form of CCBHs as a particular choice of the integration constant $\beta$, which is
related both to the curvatures of the brane and transversal section.
It is indeed this choice the one that reproduces the metric for these objects in the
original reference \cite{Banados}. We then computed thermodynamic quantities via Euclidean
methods for all dimensions higher than three. In even dimensions, the action
vanishes identically. It is therefore reassuring the fact that the Noether-Wald
charges vanish, as well, as this definition gives a zero mass at infinity and zero entropy
at $r=r_{+}$. In odd dimensions, the only
nonvanishing quantity obtained is the vacuum energy density.  As a matter of fact, this result
appears as the thermodynamic counterpart to the argument in Ref.\cite{GOP}, where the use of Noether
charges gives rise to the same expression, with all the problems of physical meaning exposed therein.\\
Following the line of reasoning in Ref.\cite{GOP}, we have shown that CCBHs are devoid
of any physical property as mass, angular momentum or entropy,  what is at odds with the interpretation of
this solution as a black hole.

\bigskip

\section{Acknowledgments}

We would like to thank I. Araya, C. Arias and P. Sundell for insightful comments. P.G. is a UNAB Ph. D. Scholarship Holder. This work was
funded in part by FONDECYT Grant No. 1170765, UNAB Grant DI-1336-16/R and CONICYT Grant DPI 20140115.

\section*{Appendices}
\appendix

\section{Kronecker delta of rank $p$} \label{Delta}

The totally-antisymmetric Kronecker delta of rank $p$ is defined as the
determinant
\begin{equation}
\delta _{\left[ \mu _{1}\cdots \mu _{p}\right] }^{\left[ \nu _{1}\cdots \nu
_{p}\right] }:=\left\vert
\begin{array}{cccc}
\delta _{\mu _{1}}^{\nu _{1}} & \delta _{\mu _{1}}^{\nu _{2}} & \cdots &
\delta _{\mu _{1}}^{\nu _{p}} \\
\delta _{\mu _{2}}^{\nu _{1}} & \delta _{\mu _{2}}^{\nu _{2}} &  & \delta
_{\mu _{2}}^{\nu _{p}} \\
\vdots &  & \ddots &  \\
\delta _{\mu _{p}}^{\nu _{1}} & \delta _{\mu _{p}}^{\nu _{2}} & \cdots &
\delta _{\mu _{p}}^{\nu _{p}}%
\end{array}%
\right\vert \,.
\end{equation}%
A contraction of $k\leq p$ indices in the Kronecker delta of rank $p$
produces a delta of rank $p-k$,
\begin{equation}
\delta _{\left[ \mu _{1}\cdots \mu _{k}\cdots \mu _{p}\right] }^{\left[ \nu
_{1}\cdots \nu _{k}\cdots \nu _{p}\right] }\,\delta _{\nu _{1}}^{\mu
_{1}}\cdots \delta _{\nu _{k}}^{\mu _{k}}=\frac{\left( N-p+k\right) !}{%
\left( N-p\right) !}\,\delta _{\left[ \mu _{k+1}\cdots \mu _{p}\right] }^{%
\left[ \nu _{k+1}\cdots \nu _{p}\right] }\,,
\end{equation}%
where $N$ is the range of indices.

\section{Curvature Tensors for the Black Brane ansatz} \label{CCBHR}

For the ansatz (\ref{ansatz}), the Riemann tensors are
\begin{eqnarray}
R_{rb}^{ra} & = & -\frac{1}{A(r)B(r)}\left (\frac{A^\prime(r)}{B(r)}\right )^{\prime}\delta^a_b\,, \label{34_1} \\
R^{rn}_{rm} & = & -\frac{1}{C(r)B(r)}\left (\frac{C^\prime(r)}{B(r)}\right )^{\prime}\delta_m^n\,, \label{34_2} \\
R_{bm}^{an} & = & -\frac{C^\prime(r)A^\prime(r)}{A(r)C(r)B^2(r)}\delta^{[an]}_{[bm]}\,, \label{34_3}\\
R^{ab}_{cd} & = & \frac{1}{A^2(r)}\mathcal{R}^{ij}_{kl}(\omega)-\left(\frac{A^\prime(r)}{A(r)B(r)} \right )^2\delta^{[ab]}_{[cd]}\, \label{34_4} \\
R_{nm}^{pq} & = & \frac{1}{C^2(r)}\mathcal{R}_{nm}^{pq}(\gamma)-\left(\frac{C^\prime(r)}{C(r)B(r)}\right )^2\delta^{[pq]}_{[nm]}\label{34_5}\,,
\end{eqnarray}
where $\mathcal{R}$ denote the Riemann tensor calculated over the indexes of the brane or the transversal section as applicable. \\

\section{Factorization of even-dimensional renormalized AdS action} \label{Beta}

The Beta function is related to the binomial coefficient by
\begin{eqnarray}
B(a,b)&=&\int_0^1dt\,t^{a-1}(1-t)^{b-1}=\frac{\Gamma(a)\Gamma(b)}{\Gamma(a+b)} \nonumber \\
\binom{p}{q}&=&\frac{\Gamma(p+1)}{\Gamma(q+1)\Gamma(p-q+1)}
\end{eqnarray}
Picking $q+1=a$ and $q-p+1=b$, we get
\begin{eqnarray}
B(q+1,p-q+1)&=&\frac{\Gamma(q+1)\Gamma(p-q+1)}{\Gamma(p+2)} \nonumber \\
 &=& \frac{1}{(p+1)\binom{p}{q}}=\int_0^1dt\,t^{q}(1-t)^{p-q}
\end{eqnarray}
The previous is helpful to express the sum  in Eq.(\ref{factorization}) as
\begin{eqnarray}
\sum_{k=0}^{n-2}(k+1)x^{n-2-k}y^k & = & \sum_{k=0}^{n-2}\frac{(n-1)\binom{n-2}{k}}{\binom{n-1}{k+1}}x^{n-2-k}y^k \nonumber \\
 & = & n(n-1)\int_0^1 dt \, t \sum_{k=0}^{n-2}\binom{n-2}{k}[(1-t)x]^{n-2-k}(ty)^k \nonumber \\
 & = & n(n-1)\int_0^1 dt\,t[(1-t)y+tx]^{n-2}
\end{eqnarray}


\end{document}